\documentclass{jpp}
\usepackage{graphicx}

\usepackage[utf8]{inputenc}
\usepackage[T1]{fontenc}
\usepackage{amsmath}
\usepackage{xcolor}  
\usepackage{graphicx}

\usepackage{hyperref}

\title{SPECTRE: A robust solver for 3D equilibria with arbitrary topology}

\author{E. Balkovic\aff{1}, J. Loizu\aff{1}, E. Lanti\aff{2}, C. Smiet\aff{1}, C. Lazzati\aff{3}, R. Ramasamy\aff{3},  A. Goodman\aff{4}, J. Geiger\aff{4}, J. P. Graves\aff{1,5} }

\affiliation{\aff{1} \'Ecole Polytechnique F\'ed\'erale de Lausanne, Swiss Plasma Center, CH-1015 Lausanne, Switzerland \\ \aff{2} Ecole Polytechnique F\'ed\'erale de Lausanne, SCITAS, CH-1015 Lausanne, Switzerland \\ \aff{3} Proxima Fusion GmbH, Flößergasse 2, Munich, 81369, Germany \\ \aff{4} Max-Planck-Institut für Plasmaphysik, 17491 Greifswald, Germany \\ \aff{5} York Plasma Institute, School of Physics, University of York, Heslington, York YO10 5DD, United Kingdom }

\newcommand{\paran}[1]{\left( #1 \right)}
\newcommand{\vect}[1]{\boldsymbol{#1}}
\newcommand{\vecdot}[0]{\boldsymbol\cdot}

\begin{document}

\maketitle
\date{}
\begin{abstract}

We present SPECTRE, a new equilibrium code based on the Multi-Region relaxed MHD model for robustly calculating 3D equilibria with general magnetic topology, allowing for flux surfaces, magnetic islands, and chaos. The code builds on a previous MRxMHD solver, SPEC, but performs significantly better thanks to a new formulation of force, the use of a stable trust-region-based least squares minimization scheme, as well as several additional features. SPECTRE is verified through application to configurations with known equilibrium solutions, in vacuum and with finite beta, both in the fixed boundary and the free boundary mode. Notably, these include vacuum equilibria of a quasi-axisymmetric (QA) device in fixed-boundary mode, and of W7-X in the free-boundary mode, along with a classical stellarator finite-beta free-boundary case. Finally, the solver is applied to a modern optimized finite-$\beta$ quasi-isodynamic (QI) configuration, where we demonstrate calculation of a strongly-shaped equilibrium with a core island which is in agreement with a HINT calculation.

\end{abstract}

\section{Introduction}

\subsection{General 3D equilibria}

Time-independent solutions to the Magnetohydrodynamic (MHD) equations satisfying $\vect{J}\times\vect{B}=\nabla p$ are essential for the description of magnetically confined fusion plasmas. These play an important role in the design of new devices, and in the operation and interpretation of experiments in existing devices. They also form the basis for predictive calculations of macroscopic plasma stability, core micro-turbulence, and power exhaust at the plasma edge. In tokamaks, where the plasma possesses axisymmetry, the magnetic field foliates a continuous set of nested flux surfaces. This fact simplifies the equilibrium problem, which then boils down to solving the well-understood hyperbolic 2D Grad-Shafranov equation \citep{gradshafranov}. However, axisymmetry does not hold in general, e.g., in stellarators or tokamaks with resonant magnetic perturbations, or in the presence of kink modes and tearing modes. 

In such a scenario, flux surfaces can be broken, leading to a magnetic field with a changed topology, featuring magnetic islands and field-line chaos. Nevertheless, the problem of finding 3D MHD equilibrium has been and is often solved using codes such as VMEC \citep{vmec} whose formulation assumes global nested flux surfaces, thus making the equilibrium problem presumably more numerically tractable. However, these solutions, also known as ideal-MHD equilibria, feature $\delta$-function current sheets on resonant rational surfaces \citep{Helander_2014}, and in the presence of pressure gradients, also singular Pfirsch-Schlüter current densities that may introduce numerical issues, as originally pointed out by \cite{grad67}. Furthermore, the solution space is not the most general one, leaving out potential equilibrium states with more general topology. 

The importance of solving general 3D equilibria has so far led to the development of alternative solvers that do not assume flux surfaces. These can be broadly separated into two categories. The first includes relaxation codes, where the plasma evolves according to perturbations that lower the plasma energy, a process that mimics initial value resistive MHD codes but does not capture the dynamics \citep{hint}. The other approach is based on the Multi-Region relaxed MHD (MRxMHD) \citep{HOLE_HUDSON_DEWAR_2006}, where the plasma equilibrium is directly found as a set of Taylor-relaxed plasma subvolumes \citep{Taylor1974}.

\subsection{MRxMHD}

MRxMHD is a variational model for describing general plasma equilibria. The idea behind it originates from the relaxation principle devised by \cite{Taylor1974}, which states that the plasma evolves in such a way that the energy is minimized, while the global helicity is conserved. In MRxMHD, the plasma is divided into subvolumes that are separated by ideal interfaces (flux surfaces with $\vect{B}\vecdot \vect{\hat{n}}=0$), and the evolution constraint is the integrated helicity in each of these volumes. The model is made precise through an energy functional that is composed of a term involving the magnetic energy $\mathcal{W}_l$ and the helicity $\mathcal{H}_l$ of each volume:
\begin{equation}
    \mathcal{F} = \mathcal{W} + \sum_l \mu_l (\mathcal{H}_l-\mathcal{H}_l^0) = \sum_l \int_{\mathcal{V}_l} dV\: \paran{\frac{B^2}{2\mu_0}+\frac{p_l}{\gamma-1}} + \mu_l \paran{ \int_{\mathcal{V}_l}  dV\:\vect{A}\vecdot\vect{B} - \mathcal{H}^0_{l}}
\end{equation}
where $\vect{A}$ is the vector potential, $\mathcal{H}^0_l$ are the conserved helicities, $\mu_l$ are Lagrange multipliers, and $p_l$ are subvolume pressures. Since MRxMHD does not rely on the strong constraint of continuous helicity conservation as in ideal MHD, magnetic reconnection in the subvolumes is allowed.

The resulting MRxMHD states are known as stepped-pressure equilibria. To find these, one looks for the states where the first variation of the functional vanishes. This condition can be translated into two equalities that are to be satisfied simultaneously: the linear force-free Beltrami field equation within the volumes, and a condition on the jump (denoted in square brackets) of the total pressure across the interfaces, referred to as the force:
\begin{gather}
    \bnabla\times\vect{B} = \mu_l \vect{B} \\
    F_l = [p + B^2/2\mu_0]_l = 0
\end{gather}
Equation 1.2 implies that each subvolume undergoes Taylor relaxation \citep{Taylor1974}, and that the pressure is constant. Additionally, equation 1.3 states that interfaces may support localized pressure gradients, with the force-balance being achieved when the total pressure on each side of an interface is equal. In the limit of a large number of subvolumes and consequently interfaces, the MRxMHD model reduces to ideal MHD as shown by \citet{infinite_vol_lim}.

\subsection{SPEC}

The Stepped Pressure Equilibrium Code (SPEC) solves equations 1.2 and 1.3 to find MRxMHD equilibria. The magnetic field is represented in terms of a vector potential $\vect{A}$ that is, in turn, expressed in a global polynomial-Fourier basis (polynomials are either Zernike in the innermost volume, or Chebyshev in the outer shells). The geometry of the interfaces is represented in cylindrical coordinates ($R, Z, \phi$) using a double Fourier series (only the stellarator-symmetric part is shown here)
\begin{gather}
    R_v (\theta, \zeta) = \sum_{m,n} R_{v,m,n} \cos(m \theta - n\zeta) \\
    Z_v (\theta, \zeta) = \sum_{m,n} Z_{v,m,n} \sin(m \theta - n\zeta)
\end{gather}
where $\theta, \zeta$ are general poloidal and toroidal angles. This representation of interface geometry is not unique unless additional angle constraints are imposed \citep{hirshman}. Alternative forms such as the Henneberg representation \citep{henn_repr} can be used, which is discussed further in Appendix A.

To solve the Beltrami field in each subvolume (Eq 1.4), SPEC requires the total toroidal and poloidal fluxes enclosed by each subvolume, $\Delta\Psi^t_l$ and $\Delta\Psi^p_l$, along with the total subvolume helicity $\mathcal{H}_l$. Furthermore, by iterating on helicity $\mathcal{H}_l$ and the poloidal flux $\Delta\Psi^p_l$, SPEC can be run with various other combinations of constraints: ($\Delta\Psi^t_l$, $\Delta\Psi^p_l$, $\mu_l$), ($\Delta\Psi^t_l$, $I^\textrm{vol}_l$, $I^\textrm{surf}_l$), or ($\Delta\Psi^t_l$, $\iota^+_l$, $\iota^-_l$), where $I^\textrm{vol}_l$ is the integrated toroidal current in each volume, $I^\textrm{surf}_l$ is the surface current on each interface \citep{Baillod_2021}, and $\iota^-_l$ and $\iota^+_l$ represent the rotational transform on each side of each interface.

While the Beltrami equation is linear and can be solved directly, the simultaneous solution of this equation and the pressure jump condition is a nonlinear problem, requiring iteration on the interface geometry. The iteration is performed until equation 1.3 is solved to a sufficient tolerance, which in SPEC is determined by looking at the Fourier spectrum of the force at each interface. The process is orchestrated by a Newton-type Powell method, which attempts to find an exact zero of the truncated force spectrum, with the truncation leading to a square system with an equal number of harmonics describing the force and the interfaces ($m_\textrm{pol}$ in poloidal and $n_\textrm{tor}$ in toroidal direction). The root-finding procedure relies on the calculation of the "force gradient", namely the derivative of force harmonics with respect to interface harmonics, which has been derived analytically \citep{Baillod_2021}.

SPEC has been used successfully for a wide range of studies. These involve the calculation of fixed-boundary \citep{Hudson2012} and free-boundary \citep{hudson2025} finite-$\beta$ equilibria in simplified 3D configurations. Furthermore, the code has enabled optimization of coils for realistic plasma equilibria with islands using the two-stage \citep{baillod_2022} and single-stage approach \citep{smiet_2025}. Additional applications worth pointing out are the study of sawteeth crashes in tokamaks \citep{zhisong2025}, the prediction of neoclassical tearing modes in slab geometry \citep{Balkovic_2025}, and the calculation of equilibrium $\beta$-limits in classical stellarators \citep{loizu_2017}, including the effects of bootstrap current \citep{Baillod_Loizu_Qu_Arbez_Graves_2023}. Despite such applications, SPEC has proven to be numerically fragile and unstable in the calculation of modern optimized stellarator configurations, which typically have strongly shaped geometries. Indeed, unless one starts with an extremely good guess for the interfaces, which somewhat defeats the purpose of using SPEC, the root-finding scheme fails to find solutions. Additionally, obtaining convergence in numerical parameters is difficult, and small parameter changes, such as adding extra harmonics in the poloidal spectrum, can lead to failure.

\section{New framework for minimization}

Inspired by the strong theoretical basis and the wide applicability of the MRxMHD model, we have developed a more robust framework for computing MRxMHD equilibria. This has resulted in the development of a new code, SPECTRE, which relies on several fundamental changes with respect to SPEC that improve robustness and flexibility.

\subsection{Improved force metric}

\begin{figure}
    \centering
    \includegraphics{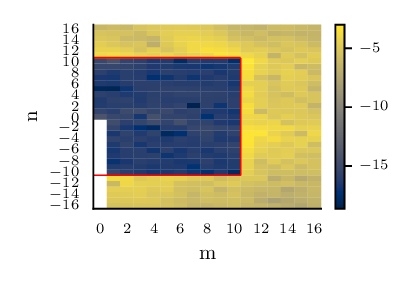}
    \vspace{-.4cm}
    \caption{Spectrum of the force for an optimized equilibrium in SPEC where only components inside the red box are targeted using root finding (data from \cite{abaillod_thesis})}
    \label{fig:fig_fmn}
\end{figure}

SPEC represents the vector potential using $m_\textrm{pol}$ poloidal and $n_\textrm{tor}$ toroidal harmonics, but the force, related to $B^2=(\nabla\times \vect{A})^2$), inherently contains a Fourier spectrum with many more harmonics. However, SPEC truncates the force to ($m_\textrm{pol}$, $n_\textrm{tor}$) harmonics, which leads to the loss of information and fictitious convergence. This seems to be the main source of the fragility of SPEC force minimization. Evidence for that can be found by analysing the residual force on an interface of a SPEC equilibrium solution. Figure \ref{fig:fig_fmn} shows an example for W7-X \citep{Wolf_2017} where the Newton method finds a zero of the force spectrum in the ($m_\textrm{pol}$, $n_\textrm{tor}$)  resolution window (modes inside the red box) but leaves a large error in the rest of the spectrum. In fact, while machine precision is achieved in the truncated Fourier space of the force, the average force residual is still around 8 kPa. Such an error also tends to cause problems for subsequent runs where one wants to increase the Fourier resolution, since in this case, the solver is given a poor initial guess with a large initial force residual. 

In SPECTRE, the force is instead represented in real space on a chosen ($\theta$, $\zeta$) grid that samples an interface, with an adjustable spacing which can be chosen appropriately to balance speed against accuracy of force representation. This representation has the benefit that all the force components have an equal scaling, something that is not true with the Fourier description of the force. Furthermore, using a general representation of the force on an arbitrary mesh allows for various advanced sampling schemes, e.g., denser sampling in the regions of high curvature on an interface. 

\begin{figure}
    \centering
    \includegraphics{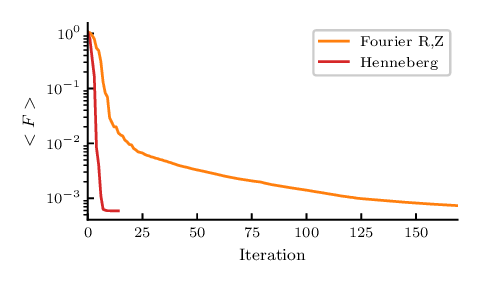}
    \vspace{-.4cm}
    \caption{Evolution of $\langle F \rangle$ during an optimization,  performed in either cylindrical Fourier $R,Z$ (orange) or Henneberg space (red). The optimization corresponds to the calculation of the QA vacuum equilibrium from section 3.2.}
    \label{fig:fig_fconvergence}
\end{figure}

\subsection{Trust-region least-squares minimization}

SPEC attempts to find an exact zero of the force-truncated spectrum, an approach that is prone to two problems. First, root-finding, such as the Powell method used in SPEC, tends to be fragile and very sensitive to initial conditions \citep{num_recipes}. Second, at any finite, practical Fourier resolution, one should not generally expect to be able to find a perfect equilibrium, and can merely expect to find a good approximation. These two issues, in addition to the problem of force representation, have led us to change the approach and use trust-region-based methods \citep{scipy, dogbox} to minimize the real-space force in order to find equilibrium states in SPECTRE. The trust-region methods are used due to their robustness, which stems from using quadratic models to locally approximate an objective function. The quadratic expansion allows the algorithms to pick both the step size and direction dynamically, while guaranteeing that the steps remain bounded to a domain. The algorithm is based on a second-order Taylor expansion of the function, but since the problem is framed in the least-squares sense, one only needs to calculate the objective function and its derivatives, without the need for the more expensive second derivatives. The force minimization target in SPECTRE is
\begin{align*}
    \min_{R_{lmn},\: Z_{lmn}} \:\sum_l \sum_{ij} F_l^2(\theta_i,\zeta_j)
\end{align*}
where the first sum is over the interfaces and the second sum is over the grid points of each interface. The force error in SPECTRE is tracked through the root-mean-square of the flux surface averages of the squared force on each interface $\langle F \rangle = \sqrt{\sum_l \langle F_l^2 \rangle / N_\textrm{vol}} $, which properly takes into account that the force sampling grid is not uniform in physical space. Convergence of a typical force minimization can be seen in Figure \ref{fig:fig_fconvergence} (orange curve), where SPECTRE starts from an initial guess with a large force error and optimizes the interface geometry until 3 orders of magnitude reduction in $\langle F \rangle$ is achieved.  

\subsection{Additional improvements}

There are a few additional features of SPECTRE that aid in the finding of MRxMHD equilibria. First, the intersection of ideal interfaces in the MRxMHD framework is not allowed, yet a numerical scheme, such as the one used in SPEC, may inadvertently end up in a state where this happens. Therefore, SPECTRE includes a check on the interface intersection, which can lead to the minimizer rejecting a given trust region if it leads to an invalid interface geometry. Another issue with finding force balance in MRxMHD is that one needs to provide an initial set of non-intersecting interfaces, something that can be surprisingly difficult in strongly shaped domains and/or where many interfaces are required. Thus, we have coupled SPECTRE to map2disc \citep{Babin_2025}, a code that solves the Laplace equation in a general disc domain. Iso-surfaces of the solution are then used as an initial guess for MRxMHD interfaces. Finally, while SPECTRE features an entirely different approach to finding force balance, the Beltrami field solver is in fact the same one as used in SPEC. Thus, all the known properties and capabilities of the SPEC Beltrami fields \citep{Hudson2012, zhisong} carry over to SPECTRE. The code is written in modern Python and links to a Fortran-based extension for calculating the Beltrami fields.

\section{Verification calculations}

Next, we present a set of test cases that are used to verify the correct calculation of equilibria. These include primarily vacuum magnetic fields for which solutions are known, and are supplemented with a finite-beta equilibrium of a rotating ellipse, along with a demonstration of a finite-beta QI optimized stellarator configuration. When not otherwise discussed, the initial interface geometry is initialized either with inward interpolation of the plasma boundary or using map2disc \citep{Babin_2025}. Such initial guesses are typically far from the converged equilibrium, but this is not an issue for SPECTRE, which has good convergence properties as will be shown.

\subsection{Axisymmetry}

\begin{figure}
    \centering
    \includegraphics[width=0.6\columnwidth]{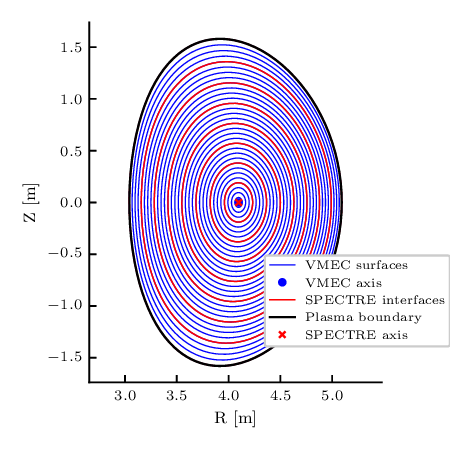}
    \vspace{-.4cm}
    \caption{The cross section of the D-shaped tokamak test case, including the VMEC flux surfaces, SPECTRE interfaces ($N_\textrm{vol}=8$), and matching magnetic axes from the two codes.}
    \label{fig:figtokamak}
\end{figure}

\begin{figure}
    \centering
    \includegraphics{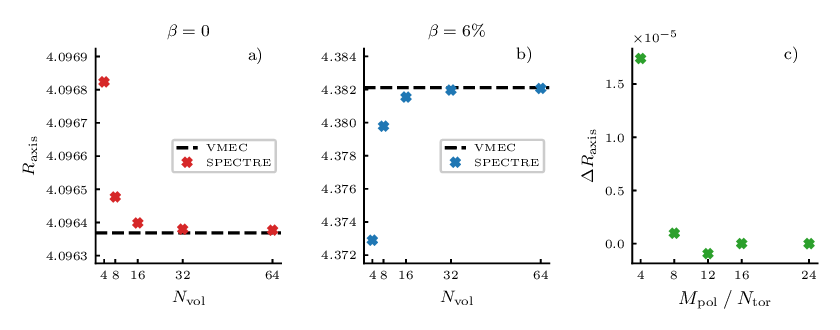}
    \vspace{-.4cm}
    \caption{Convergence of the magnetic axis position in axisymmetry as a function of $N_\textrm{vol}$ for a) $\beta=0$ and b) $\beta=6\%$. A similar scan in c), now as a function of the Fourier resolution, with the y-axis showing the difference in axis position relative to the largest resolution.}
    \label{fig:fig_tok_conv}
\end{figure}

The initial test case, coming from \citep{Hudson2012}, is for a tokamak plasma, where we anticipate nested flux surfaces everywhere and therefore expect SPECTRE and VMEC solutions to match in the limit of many volumes \citep{Hudson2012}. We impose axisymmetry ($n_\textrm{tor}=0$), scan the number of relaxed subvolumes $N_\textrm{vol}$, and study the position of the magnetic axis. The cross section of a representative equilibrium solution using VMEC and SPECTRE is depicted in Figure \ref{fig:figtokamak}. Figures \ref{fig:fig_tok_conv}a and \ref{fig:fig_tok_conv}b show results of the parameter scan for the vacuum $\beta=0$ and $\beta=6\%$ cases. It can be observed that both codes predict essentially the same axis location, with an error that is less than 1 mm even when using only $N_\textrm{vol}=4$. Additionally, Figure \ref{fig:fig_tok_conv}c shows that the magnetic axis position is also converged in terms of the Fourier resolution, and that only $m_\textrm{pol}\geq 4$ is needed. VMEC calculation was performed with $m_\textrm{pol}=32$, $N_\textrm{S}=6000$, $F_\textrm{tol}=10^{-14}$. 

\subsection{QA vacuum}

\begin{figure}
    \centering
    \includegraphics{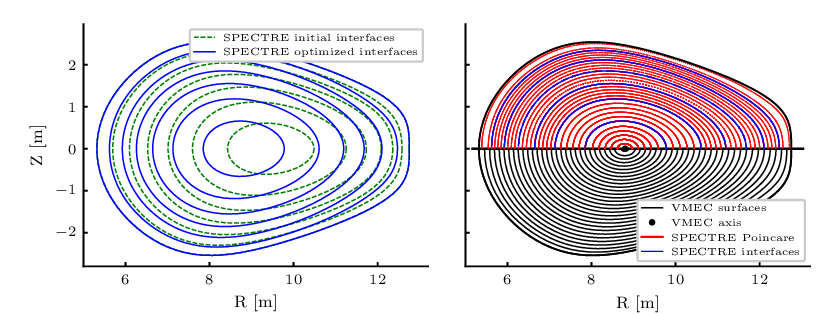}
    \vspace{-.4cm}
    \caption{a) Initial and optimized SPECTRE interfaces demonstrating robustness for a poor initial geometry guess, b) Comparison of SPECTRE interfaces and Poincaré trace with VMEC flux surfaces for the vacuum QA configuration.}
    \label{fig:fig_qavac_infin}
\end{figure}

\begin{figure}
    \centering
    \includegraphics[width=1.\textwidth]{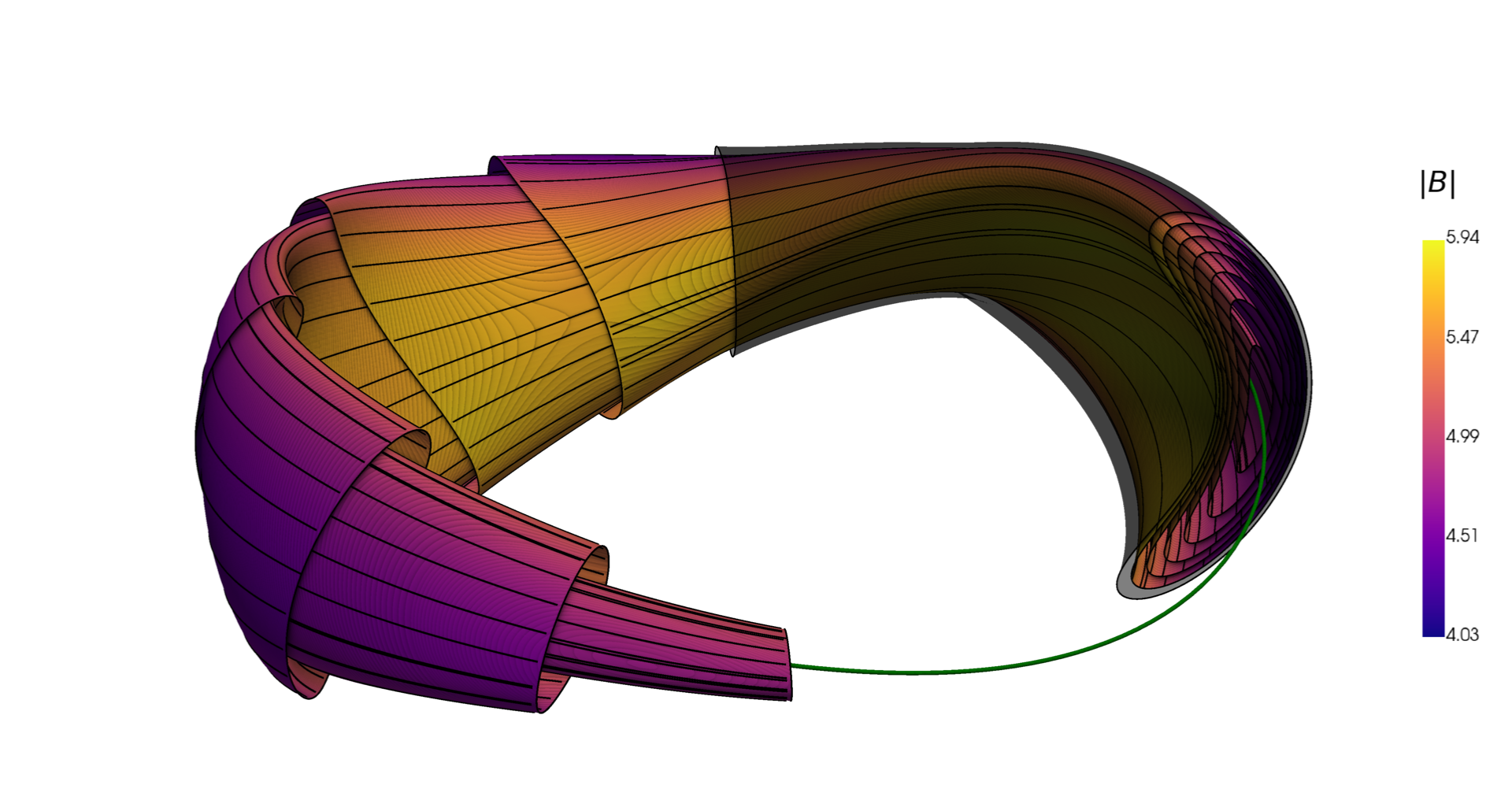}
    \vspace{-.4cm}
    \caption{A 3D depiction of the interfaces and the plasma boundary (outermost black surface) for the optimized QA vacuum equilibrium, with the magnetic axis shown in green and the color representing magnetic field strength.}
    \label{fig:fig_qa3d}
\end{figure}

The following verification case is of a QA reactor-scale configuration optimized for neoclassical confinement and stability \citep{Henneberg_2019}. It features a strongly shaped plasma boundary, with a narrow and elongated bean cross-section, and is thus a great stress-test of the robustness of the force minimization algorithm. We first look at the vacuum version of the fixed-boundary equilibrium, where the current and pressure are set to zero and the input is the plasma boundary. The initial set of interfaces created using map2disc  \citep{Babin_2025} is shown in Figure \ref{fig:fig_qavac_infin}a, along with the final set of interfaces obtained using force minimization. It is evident that the initial guess is far from the final state, demonstrating the global robustness of SPECTRE and the ability for cold-start optimization. Figure \ref{fig:fig_qavac_infin}b shows the interfaces and Poincaré trace of the final SPECTRE equilibrium, which shows agreement with VMEC flux surfaces. The final set of interfaces and the plasma boundary are shown in 3D in Figure \ref{fig:fig_qa3d}.

\begin{figure}
    \centering
    \includegraphics{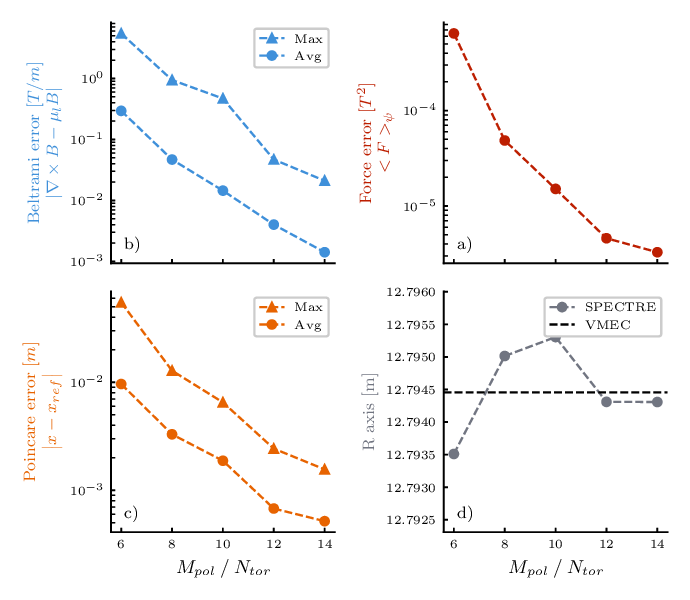}
    \vspace{-.4cm}
    \caption{Convergence trends for the QA vacuum equilibrium as a function of the Fourier resolution: a) Error in the Beltrami field (maximum of per-volume error), b) Residual $\langle F \rangle$,  c) Poincaré error between 1-volume and 7-volume cases, d) Magnetic axis position (VMEC axis shown in dashed black).}
    \label{fig:fig_qa_conv}
\end{figure}

Due to the strongly shaped boundary, one might anticipate dependence of the converged equilibrium on the Fourier resolution. The influence of resolution is studied in Figure \ref{fig:fig_qa_conv}, where the converged equilibrium is checked for a range of ($m_\textrm{pol}$, $n_\textrm{tor}$), starting at $m_\textrm{pol}=n_\textrm{tor}=6$ (resolution of the plasma boundary). Figure \ref{fig:fig_qa_conv}b shows the convergence of $\langle F \rangle$, indicating that more Fourier modes allow for better equilibria to be recovered. The achievable minimum force is directly related to the error in the Beltrami field (Figure \ref{fig:fig_qa_conv}a), which limits how accurately the force can be calculated and thus minimized. 

\begin{figure}
    \centering
    \includegraphics{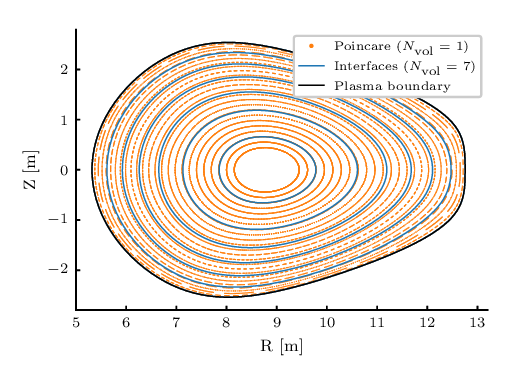}
    \vspace{-.4cm}
    \caption{Comparison of Poincaré trace for a single-volume Beltrami field and a 7-volume SPECTRE solution with interfaces in force balance.}
    \label{fig:fig_qavac_vols_1_7}
\end{figure}

This vacuum configuration, having no pressure or current, seems to have no islands and a complete nesting of flux surfaces. Thus, one would expect that the MRxMHD solution has the same global field, regardless of the number of interfaces used, as confirmed in Figure \ref{fig:fig_qavac_vols_1_7}. This fact is quantitatively verified by comparing a 1-volume solution (no interfaces, only the plasma boundary), consisting of a single Beltrami field for the entire plasma, against the 7-volume solution, requiring interface optimization. The comparison metric used here is the "Poincaré error", found by tracing the field for a single toroidal turn and calculating the relative displacement of the field between the two equilibria, a process repeated for several initial points. Figure \ref{fig:fig_qa_conv}c shows that the Poincaré error between 1-volume and 7-volume solutions also decreases with increasing Fourier resolution. Lastly, we check the position of the magnetic axis (see Figure \ref{fig:fig_qa_conv}d) and find that the consecutive SPECTRE runs with an increased number of Fourier modes converge to an axis position, which is very close to the one predicted by VMEC ($\approx$1mm). The VMEC calculation was performed with $m_\textrm{pol}=n_\textrm{tor}=22$, $N_\textrm{S}=220$, $F_\textrm{tol}=10^{-16}$. The SPECTRE force minimization required 9 cpu-hours (1.3 hrs wall clock time on 7 cpus) for the case with $m_\textrm{pol}=n_\textrm{tor}=6$ and 120 cpu-hours (8.6 hrs wall clock time on 14 cpus) in the case of $m_\textrm{pol}=n_\textrm{tor}=14$.

\subsection{Free-boundary W7X vacuum}

\begin{figure}
    \centering
    \includegraphics[width=0.97\columnwidth]{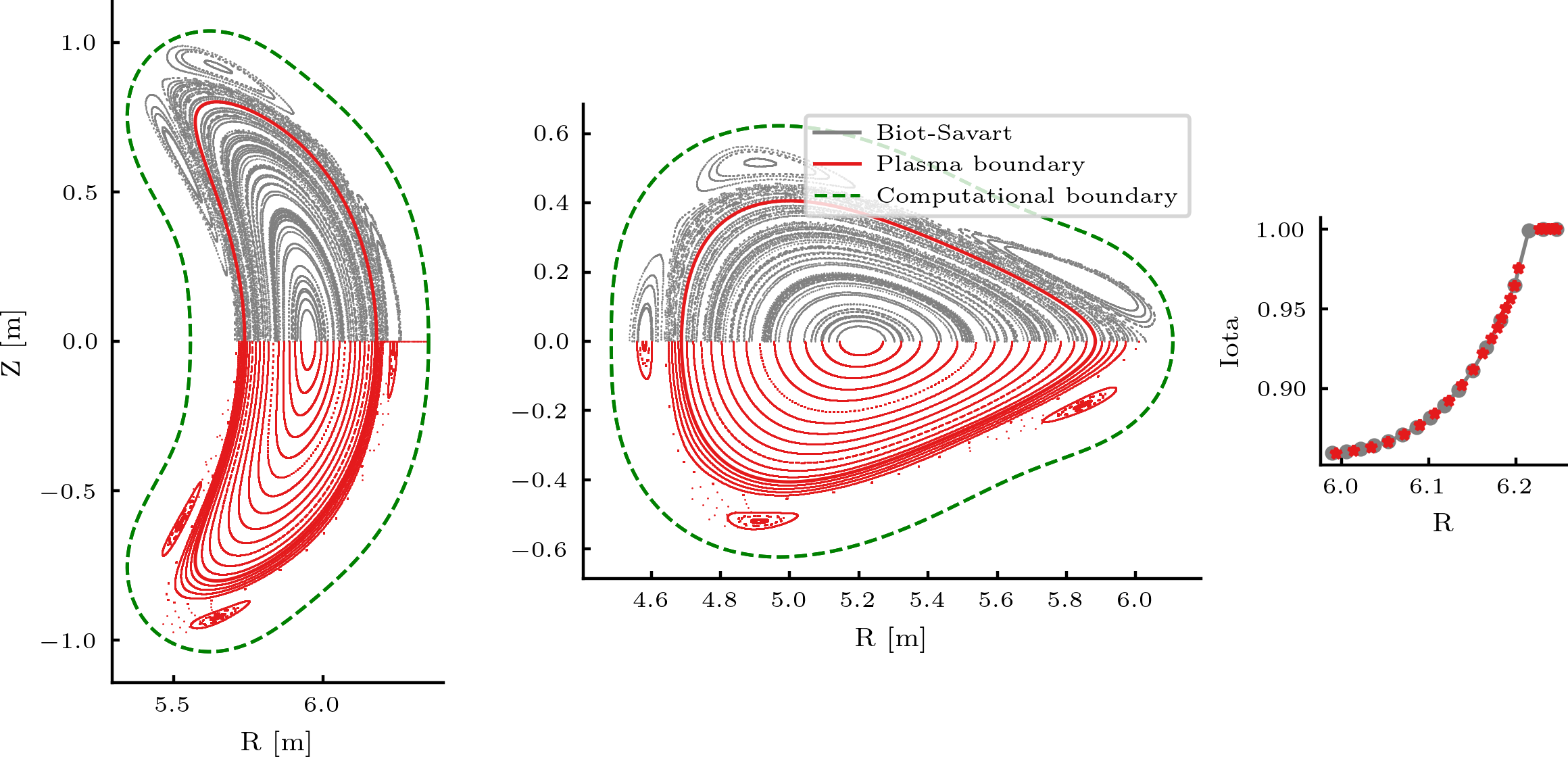}
    \vspace{-.1cm}
    \caption{Comparison of Poincaré trace from the Biot-Savart evaluation of the coils, and from the SPECTRE free-boundary solution at a) $\phi=0$ and b) $\phi=\pi/10$. Iota profile at $\theta=\phi=0$ of the two fields shown in c).}
    \label{fig:fig_w7xvac}
\end{figure}

Previous examples were done using a fixed plasma boundary that is provided at the start and fixed during minimization. However, the MRxMHD framework also naturally allows for finding free-boundary equilibria, where one solves for a plasma subject to profiles and an external field generated by coils (for more details, see \cite{Hudson_2020, hudson2025}). The free-boundary mode in SPECTRE is verified for the vacuum W7-X standard configuration OP 1.1 \citep{Wolf_2017}. The domain consists of a single toroidal plasma volume and an annular vacuum volume, solved with a different boundary condition on the outer surface $\vect{B}\vecdot \vect{\hat{n}}=\vect{B}_\textrm{coils}\vecdot \vect{\hat{n}} + \vect{B}_\textrm{plasma}\vecdot \vect{\hat{n}} \neq 0$ that is referred to as the computational boundary (dashed green line in Figure \ref{fig:fig_w7xvac}). Finding the Beltrami field in the vacuum volume requires Picard iteration \citep{hudson2025} since the Beltrami field is solved for a given normal field, which itself depends on the Beltrami field. Figures \ref{fig:fig_w7xvac}a and \ref{fig:fig_w7xvac}b show the Poincaré plots of the field from the coils, evaluated using Biot-Savart, as well as the field extracted from the SPECTRE equilibrium. There is very good agreement as SPECTRE captures the nested flux surfaces inside the plasma volume and correctly recovers the divertor island chain associated with $\iota=5/5$. The same agreement is found for the iota profiles of the two fields (Figure \ref{fig:fig_w7xvac}c). The force minimization used 30 CPU-hours (7.5 hrs wall clock time on 4 cpus), with the resolution being $m_\textrm{pol}=n_\textrm{tor}=18$.

\subsection{Free-boundary rotating ellipse}

\begin{figure}
    \centering
    \includegraphics{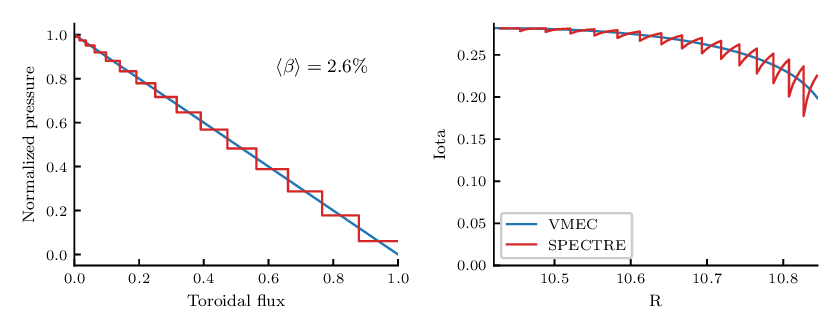}
    \vspace{-.4cm}
    \caption{a) The input pressure profiles normalized to beta on axis, b) the calculated iota profiles of the rotating ellipse equilibria in VMEC and SPECTRE.}
    \label{fig:fig_rotel_profiles}
\end{figure}

\begin{figure}
    \centering
    \includegraphics{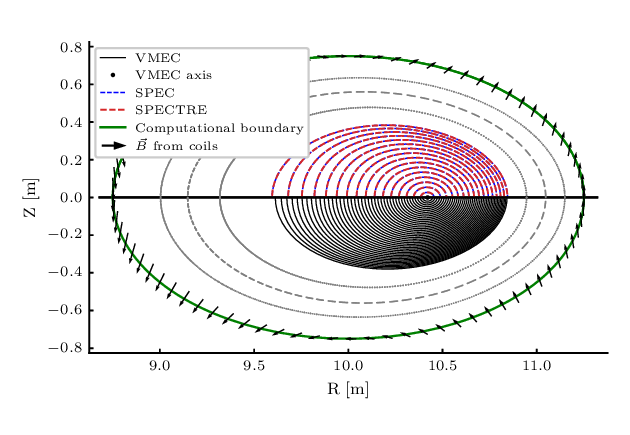}
    \vspace{-.4cm}
    \caption{The setup of the free-boundary rotating-ellipse example. Includes VMEC flux surfaces and the interfaces from SPEC and SPECTRE (exactly coincide).}
    \label{fig:fig_rotel_setup}
\end{figure}

A recent study compares VMEC and SPEC equilibria for a finite-$\beta$, zero net current (imposed directly using the current constraint \citep{Baillod_2021}) rotating ellipse configuration in the free-boundary mode with a set of external coils \citep{hudson2025}. While the flux surface geometry seems relatively simple, the equilibrium has a finite pressure profile (see Figure \ref{fig:fig_rotel_profiles}a) and a large corresponding Shafranov shift. Figure \ref{fig:fig_rotel_setup} shows the setup of the calculation, including the computational boundary and the external field, along with the converged VMEC flux surfaces, the optimized SPEC interfaces \citep{hudson2025}, and the optimized SPECTRE interfaces ($N_\textrm{vol}=16$). The three codes find nearly identical fields, as indicated by the match in the location of the magnetic axis: $R_\textrm{VMEC}=10.4205 \textrm{m}$, $R_\textrm{SPEC}=10.4197 \textrm{m}$, $R_\textrm{SPECTRE}=10.4196 \textrm{m}$. We remark that SPECTRE is run independently from VMEC and SPEC, namely the initial interface guesses are generated without relying on the solutions of the other two codes. This is reflected in the initial axis which has $R_\textrm{SPECTRE, init}=10.4 \textrm{m}$. The iota profile extracted from the SPECTRE equilibrium is the same as in SPEC, and this approaches the iota profile from VMEC (see Figure \ref{fig:fig_rotel_profiles}b) as more interfaces are used. The force minimization required around 10,000 cpu-hours (117 hrs wall clock time on 85 cpus) with the resolution being $m_\textrm{pol}=n_\textrm{tor}=12$.

\subsection{Finite-$\beta$ QI equilibrium}

To demonstrate the application of SPECTRE to modern stellarators with strong shaping, we present the equilibrium calculation of a candidate next-step 4-period QI device. We study this configuration with a given pressure profile ($\langle \beta \rangle_\textrm{volume}=1.5\%,\: \beta_\textrm{axis}=3\%$) and zero net toroidal current (which is characteristic of QI configurations). Due to the lack of a good reference solution such as in vacuum equilibria, the goal of this study is to compare SPECTRE to an alternative 3D equilibrium solver used in the community, HINT \citep{hint}. HINT is an iterative code that finds equilibria through plasma relaxation, achieved through alternating adjustment of the magnetic field and pressure. While HINT was run in the free-boundary mode consistent with coils, SPECTRE is used here in the fixed-boundary mode, employing an approximate HINT core flux-surface as the plasma boundary. 

\begin{figure}
    \centering
    \includegraphics[width=1.00\columnwidth]{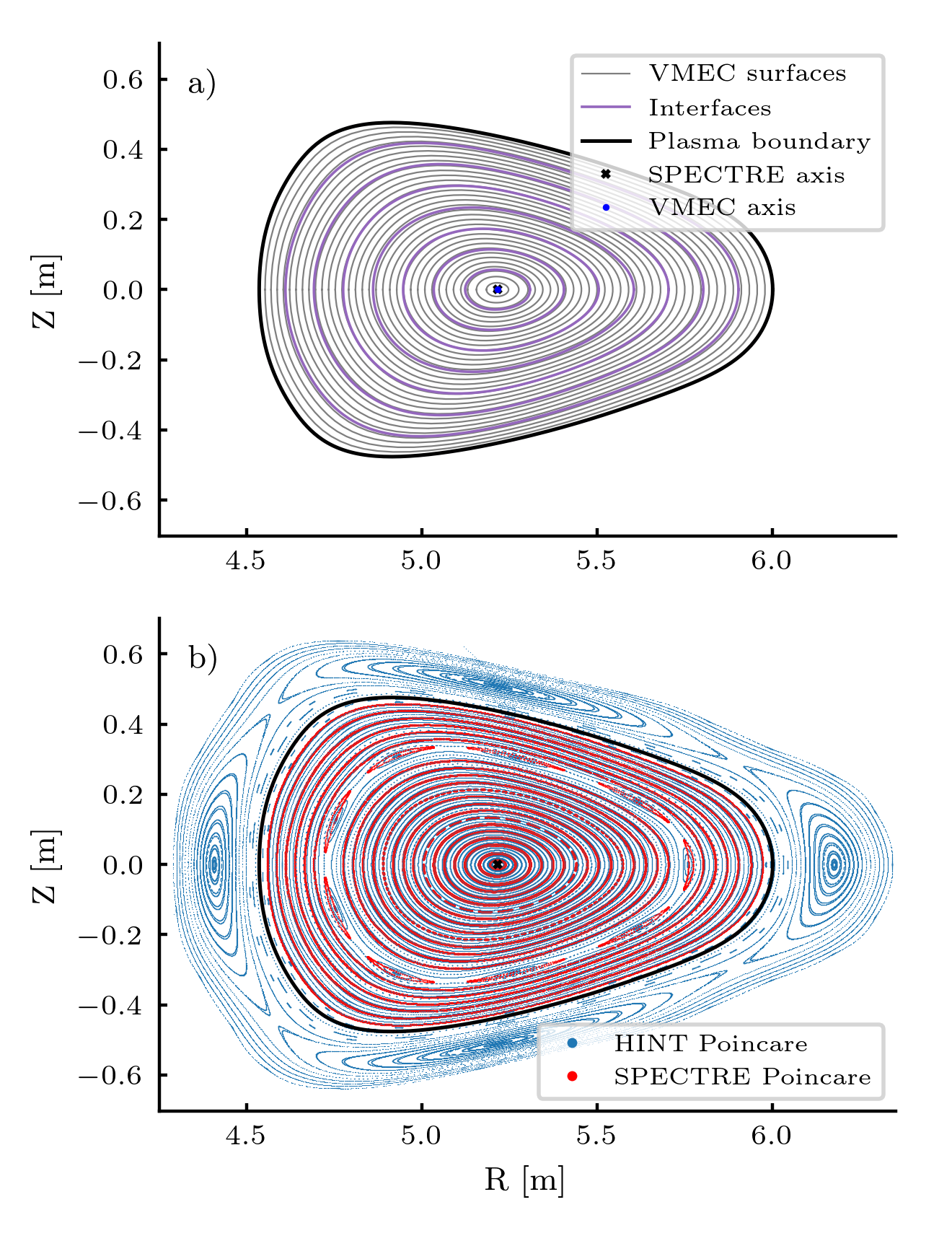}
    \vspace{-.4cm}
    \caption{Calculated QI finite-$\beta$ equilibria. a) Comparison of VMEC flux surfaces and SPECTRE interfaces b) Poincaré traces of the equilibrium field from SPECTRE (red) and HINT (blue), both depicting the 8/9 core island chain. The black line is the fixed plasma boundary used in VMEC and SPECTRE.}
    \label{fig:fig_qicomparison}
\end{figure}

\begin{figure}
    \centering
    \includegraphics[width=0.6\columnwidth]{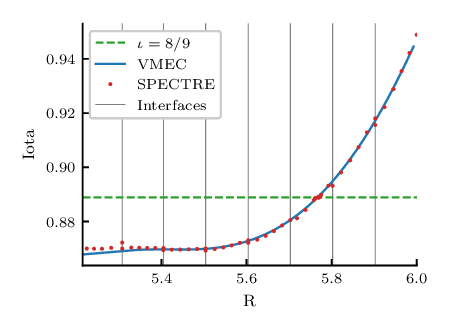}
    \vspace{-.4cm}
    \caption{Rotational transform profiles of the QI configuration found using VMEC (blue line) and SPECTRE (red points). SPECTRE iota shows flattening around the $\iota=8/9$ resonance where the island exists.}
    \label{fig:fig_qiiota}
\end{figure}

Using the extracted boundary and the profiles, we run SPECTRE force minimization with $N_\textrm{vol}=8$ relaxed subvolumes to find an equilibrium state, interfaces of which are shown in Figure \ref{fig:fig_qicomparison}a. We additionally run fixed-boundary VMEC \citep{vmec} and obtain flux surfaces which generally match with SPECTRE interfaces. The magnetic axis in the two codes is similar, with an error less than $\Delta R=2\times10^{-4}\textrm{m}$. Figure \ref{fig:fig_qiiota} depicts the rotational transform profiles for SPECTRE and VMEC, which also show agreement.

Next, we compare the magnetic field of the SPECTRE equilibrium, calculated only inside the boundary, with the HINT field, as shown in Figure \ref{fig:fig_qicomparison}b. Despite a substantially different physical model and numerical approach, SPECTRE and HINT show a good qualitative agreement in the Poincaré trace. In addition to the match in the location of the magnetic axis, $\Delta R=2\times10^{-3}\textrm{m}$, both codes predict the existence of an island chain at the resonant surface $\iota=8/9$ and flux surfaces filling the rest of the volume inside the plasma boundary. The island found through the two codes has the same phase, a similar position of the O-point, $\Delta R=7\times10^{-3}\textrm{m}$ at $\phi=\pi/4$, and a similar island width. The iota profile in SPECTRE, as shown in Figure \ref{fig:fig_qiiota}, shows small flattening around the island O-point with the correct resonance.

In this example calculation, the interfaces in SPECTRE are parametrized quadratically in toroidal flux $\Psi^t_l$, leading to a linear spacing in an approximate minor radius $r\approx\sqrt{\Psi^t_l}$. Choosing a good interface parametrization is an important aspect of running SPECTRE, as one wants to avoid interfaces close to separatrices of islands and close to low-order rational surfaces, neither of which is entirely known a priori. The choice of quadratic flux spacing seems to have satisfied those conditions here, and we have checked that perturbing this spacing does not impact the overall equilibrium. However, more work needs to be done to test the exact dependence of island properties on the interface parametrization, as well as finding a good scheme for interface placement. Finally, in this calculation, the Fourier resolution is set to $m_\textrm{pol}=12$, $n_\textrm{tor}=11$, with higher resolution leading to no discernible change in the islands and the equilibrium field.

The work presented for the optimized QI stellarator is a first-of-a-kind qualitative comparison of SPECTRE with HINT, using a fixed plasma boundary in SPECTRE which only approximates a HINT flux surface. A more rigorous study will follow in the future, operating SPECTRE also in the free-boundary mode directly from the coils. Such a calculation will enable a quantitative comparison of the two codes, including a detailed study of core islands as well as the divertor island chain in the edge.

\section{Discussion and outlook}

We presented the development of SPECTRE, a new code for finding equilibrium states of the MRxMHD model \citep{HOLE_HUDSON_DEWAR_2006}  with good numerical stability and robustness. The solver relies on an improved representation of the force in real space, which is minimized in a least-squares sense using a well-behaved trust-region method \citep{dogbox}. A range of equilibria was studied to demonstrate the improved behavior compared to the approach in SPEC \citep{Hudson2012}. These have shown that we can recover correct fixed-boundary equilibria with fully nested flux surfaces and without islands, in which case SPECTRE solutions agree with those from VMEC. The match was also established for free-boundary equilibria, correctly predicting the W7-X vacuum divertor island chain and a finite-$\beta$ rotating ellipse equilibrium. We have furthermore indicated that the error in the calculated equilibria converges with Fourier resolution. Finally, SPECTRE was applied to a finite-$\beta$ optimized QI candidate configuration. Here, SPECTRE calculated an equilibrium with a core island chain that has also been found using HINT \citep{hint}. Further validation still remains to be done, in particular with regard to the impact of interface placement on the equilibrium, as well as the calculation of the true free-boundary equilibrium and a comparison with HINT. More generally, we foresee further benchmarking studies comparing SPECTRE to HINT in other relevant stellarator concepts, as well as comparisons to initial-value resistive MHD codes such as JOREK \citep{Hoelzl_2021} and M3D-C1 \citep{m3dc1}. We hope that SPECTRE can help bring about the true potential of MRxMHD and become a useful tool for stellarator optimization and divertor studies. As shown in recent work, MRxMHD is also applicable in the study of nonlinear MHD stability, having been used to find saturation of classical and neoclassical tearing modes (NTM) in simple geometries \citep{Loizu_Bonfiglio_2023, Balkovic_2025}. We believe that the work done on SPECTRE will help improve upon these studies and enable the prediction of toroidal NTMs.

\section{Code availability}

The SPECTRE code is publicly available at \url{https://gitlab.com/spectre-eq/spectre} and is released under the MIT license. We welcome new users and contributions from the community.

\section{Author contributions}

E.B. and J.L., with the support of E.L., C.S., and J.P.G., conceptualized and developed SPECTRE, calculated the test equilibria, and drafted the manuscript. The optimized QI configuration was originally developed by A.G., with C.L. and J.G. leading the HINT calculations and R.R. facilitating HINT-SPECTRE comparison. All authors contributed to the discussion of the results and provided input on the manuscript.

\section{Acknowledgments}

The work on the QI finite-$\beta$ configuration was done with funding from Proxima Fusion, whom we also thank for support in providing the necessary files. We also thank Stuart Hudson for insightful discussions and for providing inputs for the free-boundary rotating ellipse test case, and Sophia Henneberg for providing files for the QA configuration.

This work was supported in part by the Swiss National Science Foundation. And by a grant from the Simons Foundation (1013657, JL). This work has been carried out within the framework of the EUROfusion Consortium, via the Euratom Research and Training Programme (Grant Agreement No 101052200 — EUROfusion) and funded by the Swiss State Secretariat for Education, Research and Innovation (SERI). Views and opinions expressed are however those of the author(s) only and do not necessarily reflect those of the European Union, the European Commission, or SERI. Neither the European Union nor the European Commission nor SERI can be held responsible for them.

\appendix

\section{Interface representation}

Representing 3D toroidal geometries using cylindrical R, Z coordinates expressed in the Fourier basis (equations 1.4 and 1.5) is a common approach used throughout stellarator codes. Such a description does not uniquely specify the angle $\theta$, which has led to the development of spectral condensation, a metric which can be minimized to constrain the representation \citep{hirshman}. SPECTRE does not inherently require such a trick, as the trust-region algorithm is still able to minimize the force. However, we have found that the null-space presented by the standard representation still leads to slow convergence and small trust-region sizes. This can be seen in Figure \ref{fig:fig_fconvergence}, where the force reduces quickly at first, but then descends slowly, requiring many iterations. 

The problem of slow convergence has been addressed by employing the Henneberg representation \citep{henn_repr}. This basis defines a unique geometrical angle (optimal in rotating ellipse geometries) and essentially defines a parameter akin to the minor radius, which is then expanded in a Fourier basis. In the same Figure \ref{fig:fig_fconvergence}, we also include a force-iteration curve for a case where the same configuration is minimized using Henneberg degrees of freedom, which now takes just $\approx10$ iterations to find an identical final state. SPECTRE is flexible since the force can be optimized in any representation, as long as a smooth and invertible mapping exists between the representation and the standard R, Z Fourier basis, in which the Beltrami field is solved. Thus, one future improvement may be to attempt minimization in the non-geometrical equal-arclength angle, which automatically has a minimized spectrum (order p=2) as shown by \cite{hirshman}. This may improve flexibility for cross-sections where a Henneberg angle is difficult to define, and may lead to a representation with a narrower Fourier spectrum.

\bibliographystyle{jpp}

\bibliography{main}

\end{document}